\documentclass[prb,amsmath,amssymb,preprint]{revtex4-1}
\usepackage{graphicx}
\usepackage{bm}
\usepackage{color,graphics}
\begin{document}

\title{Why is the earth not burning ? The earth radiative energy balance.}

\author{A. Bret}
\affiliation{ETSI Industriales, Universidad de Castilla-La Mancha, 13071 Ciudad Real, Spain}

\begin{abstract}
The concept of energy balance is a key one in climate science. Yet, students may find it counter-intuitive: while it is obvious that some energy comes in from the sun, the part coming out is more elusive. Asking them why the earth is not burning after billions of years of exposure to the sun, takes them to the question ``where does the energy goes ?'' A series of Fermi like calculations then convinces them that storage capabilities are negligible compared to the amount of energy coming in: the earth necessarily re-emits what it receives.
\end{abstract}

%\pacs{52.35.Qz, 41.75.-i, 03.65.-w}

\maketitle

\section{Introduction}
A key concept if climate science is the one of energy balance \cite{climateprimer}. The mean temperature of every planets of the solar system, including the earth, can be evaluated writing that  it eventually re-emits what it receives from the sun. Yet, the very concept of energy balance can be counter intuitive for students. While the incoming part of the balance (what comes from the sun) is obvious to them, the outgoing part is not so clear. Could it be that some energy gets stored, in some way? Curiously, that is the big part of the misunderstanding, as they don't tend to assume that the earth can loose \emph{more} than it gains. This article exposes a series of questions helping the students to make their way through this issue, and become convinced that indeed, the earth must re-emit what it receives.

As a starting point, albedo and greenhouse effects are not considered. Once the very principle of the energy balance have been established, they obviously constitute  the next steps towards a more realistic model \cite{knoxa,wolbarst}.

\section{Why the earth is not burning ?}
A one meter square surface oriented towards the sun and placed at the earth orbit receives $C_S=1370$ W, the so-called ``solar constant''. The earth radius $R_e$ is 6400 km, so that it intercepts the total solar power,
\begin{equation}\label{eq:sol_power}
    P_S=\pi R_e^2 C = 1.76\times 10^{17}~W.
\end{equation}
In just one second, the earth receives from the sun $1.76\times 10^{17}$~Joules. And it has been like this for some 4.5 billions years. How is it that the earth is not burning ? How is it that the earth has not been ``vaporized'' ? Even leaving aside the albedo, and the long or short term variations of the solar constant \cite{ipcc},  the amount of energy received during such an enormous amount of time is staggering. We find here an instance where the law of energy conservation becomes tangible and yet, enigmatic. It is obvious to everyone that energy arrives every second from the sun. If energy conservation is true, where does it go ?

Once $P_S$ has been computed, and the question asked, students's tentative solutions to the problem usually fall in two categories: 1/ Part of the energy must leave the earth (at this stage, the ``how'' is not relevant yet). 2/ Part of the energy must be stored somehow on earth. The next questions are first oriented towards the second option. Once students are convinced that the amount of energy that can be stored is negligible compared to the amount poured in, the energy balance is inescapable.

\section{Can the earth system store some energy ?}
The problem examined here is therefore: to which extent could the earth system store the energy received from the sun? Here again, students are not short of ideas and usually come up with energy storage in the biomass (vegetation and oil), or in  water under the form of heat. On a short time scale, it takes energy, through photosynthesis, to grow plants and trees, so that they can be considered as energy reservoir. On a much larger time scale, oil comes from dead organisms, and its energy content can equally be traced back to the sun. We thus now quantify, even loosely, the capacity of these reservoirs to see if they may have played a role absorbing the sun energy received in the last billion years.

\subsection{Capacity of forests as a reservoir}
One meter square of earth receives on average $P_S/4\pi R_e^2=C_S/4=342.5$ W. Considering photosynthesis can exploit some 5\% of it, a growing forest can absorb 17 W/m$^2$. Assuming a forest reaches maturity after 100 years, it will eventually contain at most $17\times 3600 \times24\times 365 \times 100=53$ GJ/m$^2$.

Let us now assume (certainly exaggerating) that 50\% of dry land is made of forest. Since dry land is 30\% of the earth surface, forests amount for some 77 10$^6$ km$^2$. Multiplying this surface by the energy density previously derived, we come to the conclusion that the amount of energy stored in vegetation is about,
\begin{equation}\label{eq:E_v}
    E_V=53~\mathrm{GJ/m}^2\times 77\times 10^6~\mathrm{km}^2=4.16\times10^{24}~\mathrm{J}.
\end{equation}
It is now worth comparing this number to the total power received from the sun derived in Eq. (\ref{eq:sol_power}). Dividing the former by the later, one finds the sun irradiance ``fills'' the vegetation reservoir is just 273 days! Because we are talking billion years, this reservoir is obviously inadequate. Let us now turn to the energy oil can store.

\subsection{Capacity of oil as a reservoir}
This calculation can obviously be made for gas, or coal, as the result will prove this kind of reservoir equally inefficient to avoid the energy balance. The conclusion is here even more straightforward. According to the most optimistic estimates of the oil industry, the total amount of conventional oil ever present in the ground was about $3.10^{12}$ barrels, of which some $1.10^{12}$ have already been extracted \cite{ukpeak}. With a 160 liters barrel, and considering a density of $\sim 1$ g/cm$^3$, this amounts to $4.8\times 10^{14}$ kg of oil. Since the combustion of 1 kg of oil releases some 42 MJ, the overall reservoir contains,
\begin{equation}\label{eq:E_o}
    E_O=2.16\times 10^{22}~\mathrm{J}~=~34~\mathrm{hours~of~sun}.
\end{equation}
Here again, we find the reservoir far too small to play a role in containing the sun energy received by the planet during the last billion years. The number itself show that adding unconventional oil, coal or gas to the calculation cannot help solve the problem.

\subsection{Capacity of the oceans as a reservoir}
Let us now check how much energy can be stored in the oceans. The volume of water they contain is about $1.3\times 10^9$ km$^3$. Considering a heat capacity of 4.1 kJ/K/kg, we find the energy absorbed by the oceans when gaining 1 degree Kelvin,
\begin{equation}\label{eq:E_w}
    E_W=5.43\times 10^{24}~\mathrm{J}~=~356~\mathrm{days~of~sun}.
\end{equation}
This third reservoir is equally found wanting. Even if an increase of 1°K degree can capture almost one year of sun energy, the process cannot proceed for more than 100 years.

These three Fermi-like calculations make it clear that the power from the sun cannot be stored on earth on a billion time scale. Indeed, we just checked that given its consequences, the imbalance time in 4.5 billions years must be smaller than $\sim 100$ hours, which amounts from some 10$^{-12}$ of the total. The image of a parking facility is here very useful to students: some 10$^{12}$ cars (the energy) want to enter the parking, but there is only one single place. The only solution is that the number of cars leaving the park in (almost) any amount of time, is exactly the number of cars entering it.

\section{Conclusion: The energy balance}
At this stage, the conclusion is inescapable: if there are no long term storage solutions, the energy coming in must eventually leave the earth. Since it cannot be evacuated to the interplanetary space through convection nor conduction, the only way out is radiation. Assuming an average temperature $T_e$, this allows to derive directly the relation,
\begin{equation}\label{eq:balance}
    \sigma T_e^4=C_S/4.
\end{equation}
Albedo and greenhouse parameters can thus be explained in the sequel.

Although the evaluation of the reservoirs capacity is highly approximate, the orders of magnitudes involved are so disparate that there is no way to escape the need for an energy balance. Accounting for unconventional oil reserves such as ``tar sands'', or for an higher photosynthesis yield, cannot yield some reservoir large enough to retain a significative portion of the energy poured by the sun over billions of years.

\begin{acknowledgments}
This work has been  achieved under projects ENE2009-09276 of the
Spanish Ministerio de Educaci\'{o}n y Ciencia and PAI08-0182-3162 of
the Consejer\'{i}a de Educaci\'{o}n y Ciencia de la Junta de
Comunidades de Castilla-La Mancha.
\end{acknowledgments}

\bibliography{BibBret}

\end{document}